\documentclass[10pt,aps,prl,twocolumn,floatfix]{revtex4-2}
\usepackage{amssymb}
\usepackage{amsmath}
\usepackage{amsfonts}
\usepackage{graphicx}

\renewcommand{\vec}[1]{\boldsymbol{#1}}
\renewcommand{\vr}{\vec{r}}
\newcommand{\vu}{{\bf u}}
\newcommand{\vx}{{\mathbf{x}}}
\newcommand{\calC}{{\mathcal{C}}}
\newcommand{\tcalC}{\widetilde{{\mathcal{C}}}}
\newcommand{\tgm}{\tilde{\gamma}_{2}}
\newcommand{\bL}{\tilde{\boldsymbol{\Lambda}}}
\newcommand{\tlambda}{\tilde{\lambda}}

\begin{document}

\title{Quantum‑Information Measure of Electron Localization}
\author{Stefano Pittalis}
\affiliation{Istituto Nanoscienze – CNR, S3, Via Campi 213A, I-41125 Modena, Italy}
\email{stefano.pittalis@cnr.it}
\author{Filippo Troiani}
\affiliation{Istituto Nanoscienze – CNR, S3, Via Campi 213A, I-41125 Modena, Italy}
\author{Celestino Angeli}
\affiliation{Dipartimento di Scienze Chimiche, Farmaceutiche ed Agrarie, Università di Ferrara, I-44121 Ferrara, Italy}
\author{Irene D'Amico}
\affiliation{School of Physics, Engineering and Technology, University of York, YO10~5DD York, United Kingdom}
\affiliation{York Center for Quantum Technologies, University of York, YO10 5DD, York, United Kingdom}
\author{Tim Gould}
\affiliation{Quantum and Advanced Technology Research Institute,
  Griffith University, Nathan, Qld 4111, Australia}
\begin{abstract}
Understanding electron localization in molecules and materials plays a central role in electronic structure theory---and will increase in importance with the rise of data-driven approaches.
The electron localization function (ELF) is widely used to visualize electron organization in molecules and materials, and it remains a central ingredient in modern density‑functional approximations. Yet its formulation retains highly empirical elements. Here we introduce a quantum-information measure of electron localization derived from the concurrence of a correlated two‑spin mixed state. This construction yields a genuine two‑point localization indicator grounded in quantum‑information theory, avoiding the heuristic normalization and chosen nonlinear remapping of the ELF. We show that atomic shells, covalent and ionic bonds, lone pairs, molecular dissociation, and charge‑transfer processes are captured. The method is straightforward to evaluate numerically.
\end{abstract}

\maketitle

{\it Introduction.}
The term {\it localization} carries many different meanings across physics—from Anderson localization in disordered systems to Wannier localization in periodic solids and spatial confinement in quantum dots. Here we focus on a distinct notion: the propensity of electrons to stay together or apart, as encoded in two‑particle correlations within regions associated with bonds, atomic shells, and lone pairs. 

Real‑space indicators of electron localization can be viewed as modern counterparts to Lewis structures \cite{Lewis16} to understand the organization of electrons in atoms, molecules, and materials. They are also pervasive in both density‑functional and wave‑function electronic‑structure methods. Among them, the electron localization function (ELF) is arguably the most widely used. Initially introduced by Becke and Edgecombe through the curvature of the like‑spin conditional pair density at the Hartree–Fock level~\cite{Becke90}, and later reinterpreted as a Pauli (Kohn–Sham) kinetic‑energy excess relative to its bosonic von Weizsäcker limit~\cite{Savin92,Savin94}, the ELF has proved invaluable for visualizing bonds and shells. 

Yet, despite its success~\cite{BMG05,RasanenCastro2008,Furness2016,Desmarais2024}, its construction includes empirical ingredients, and its semi‑local form~\cite{fnote_semi} can limit its ability to capture intrinsically nonlocal electronic features \cite{Giner2016}. This empirical content is often accepted on the grounds that ELF analyses are largely qualitative. However, ELF is also widely used to carefully analyze and compare bonds across different systems and environments. Elements of the ELF also enter modern density‑functional approximations \cite{SCAN, CHEMSCAN, LAK}, further motivating the need for a {more transparent theoretical foundation} to support quantitative objectives. Interpretative challenges for the ELF have been reported as well~\cite{SavinTheo05}. Significant efforts have been made over the years to place the ELF on safer theoretical grounds~\cite{NalewajskiELF, DobsonELF, BurdettMcCormickELF}, and Kohout's ELI family of descriptors~\cite{Kohout2004} avoids ad-hoc scaling and normalization by adopting a space-partitioning approach.

Quantum‑information approaches have meanwhile shown that electronic entanglement is a powerful descriptor of bonding, reactivity, and correlation in molecules~\cite{Boguslawski2012,Boguslawski2013,MolinaEspiritu2015,Duperrouzel2015,Kurashige2013,Stein2016,Tenti2016,Ding2021,Materia2024}. Yet most of these measures are orbital‑resolved and can vary under orbital rotations~\cite{Stein2017,Angeli2024}, limiting their suitability as universal real-space indicators. 
 
Along parallel developments, spin entanglement has been investigated in  the non‑interacting electron gas~\cite{Vedral03a,Oh04,Pittalis15a,Troiani2025a}, suggesting that a real‑space, basis‑independent localization indicator may be derived from spin entanglement~\cite{Pittalis15a}. This viewpoint offered an alternative rationalization of the internal workings of the ELF. However, the same analysis did {\it not}  overcome the empirical elements in the ELF construction.

In the following, we address the longstanding challenge of formulating electron localization {through a well-defined spin-entanglement indicator}. We achieve this by employing a measure of entanglement derived from the underlying two‑spin correlations. For each pair of spatial points, we construct a properly normalized mixed two‑spin state and quantify its entanglement via the concurrence function~\cite{Wootters1998}. As we show below, this procedure provides a physically grounded measure of localization in correlated many‑electron states and can be applied straightforwardly, yielding consistent and reliable results.

{\it Formulation.}~
Given an $N$‑electron state $\Psi$, the corresponding one‑body reduced density matrix (1RDM) is
\begin{align}
\gamma_1(\vx_1;\vx_1')
=  &N \!\! \int \! d\vx_{2}\dots \! \int \! d\vx_{N}
 \nonumber \\
&\Psi(\vx_1,\vx_2 \dots \vx_N) 
\Psi^*(\vx_1',\vx_2 \dots \vx_N)\,
\label{eqn:1RDM}
\end{align}
with natural orbitals $\psi_i$ and occupations $0\le n_i\le 1$ defined by
\begin{align}
\int d\vx_{1}'\,\gamma_1(\vx_1;\vx_1')\,\psi_i(\vx_1')
= n_i\,\psi_i(\vx_1).
\label{nt}
\end{align}
From $\gamma_1$, we may form the anti-symmetrized two-body correlation function
\begin{align}
\tgm(\vx_1,\vx_2;\vx_1',\vx_2')
= \left[\gamma_{1} \wedge \gamma_{1}\right](\vx_1,\vx_2;\vx_1',\vx_2') 
\label{eqn:dm2dm1}
\end{align}
where ``$\wedge$'' stands for the wedge product \footnote{The wedge product of the one-body reduced density matrix is defined as $\left[\gamma_{1} \wedge \gamma_{1}\right](\mathbf{x}_1,\mathbf{x}_2;\mathbf{x}_1',\mathbf{x}_2') = \gamma_{1}(\mathbf{x}_1;\mathbf{x}_1')\,\gamma_{1}(\mathbf{x}_2;\mathbf{x}_2')
- \gamma_{1}(\mathbf{x}_1;\mathbf{x}_2')\,\gamma_{1}(\mathbf{x}_2;\mathbf{x}_1')$.}.

The correlation function $\tgm$ is significant for three main reasons: (i) it  naturally generalizes the conventional Hartree and Fock terms to the correlated $\gamma_1$ \cite{Lieb83};
(ii) {as the seminal works by Bader~\cite{BaderXhole1,BaderXhole2}, Becke~\cite{Becke90}, Savin~\cite{Savin92}, and collaborators have demonstrated, $\tgm$} encodes key elements of the physics that underlies electron localization;
{and  (iii) the auxiliary object $\tgm$ can encode singlet--triplet mixing that is not visible when the concurrence is extracted directly from the pure-state 2RDM $\gamma_2$}~\footnote{We remind that $\gamma_2(\vx_1,\vx_2;\vx_1',\vx_2') \!\!\!\!\! = \!\!\!\!\!N(N-1) \!  \int \! d\vx_{3}\dots \int \! d\vx_{N} \Psi(\vx_1,\vx_2,\vx_3\dots\vx_N) \Psi^*(\vx_1',\vx_2',\vx_3\dots\vx_N)$}.

{Here we exploit the singlet--triplet mixing in $\tilde\gamma_2$ to introduce a measure of electron localization based on spin entanglement.}
{The underlying physical picture is straightforward: when two electrons share a sufficiently small region of space, the antisymmetry of the total state forces them into a singlet---and thus maximally entangled---spin configuration. 
As the two electrons move apart, the weights of the triplet components tend to increase, and the two-spin state becomes less entangled. Thus, $\tilde\gamma_2$ provides a connection between electron localization and spin entanglement that can be used to analyze both.
}

{We stress that $\tgm$ is not used here as an approximation to $\gamma_2$, but as an auxiliary quantity: $\gamma_2$ and $\tgm$ encode different information, with $\gamma_2$ determining the two-particle probability of the state under consideration and $\tgm$ carrying the local singlet--triplet structure relevant for the localization analysis developed here.} 
{We also note that the triplet (antisymmetric) component of the 2RDM has been used previously in localization-related analyses~\cite{Kohout2008TCA}}. 
The focus is thus on the state described by $\tgm$, while the full two-body cumulant \cite{KutzelniggMukherjee1999}---i.e., $\left(\gamma_2 - \tgm\right)$---is neither required nor employed. 

The properties of $\tilde\gamma_2$ deserve further comments. When $\gamma_1$ is idempotent, as in Restricted Hartree-Fock (RHF)\cite{BookHF}, $\tilde\gamma_2$ coincides with $\gamma_2$, while when $\gamma_1$ carries fractional occupation numbers---as for a multi-configurational state---$\tilde\gamma_2$ {need not satisfy pure-state $N$-representability conditions}.
 In order to stress this point and avoid any confusion, in this paper $\gamma_2$ denotes the pure-state 2RDM defined above.

To estimate the entanglement of the spin pairs, 
we consider the spatially-dependent $4\times4$ spin matrix $\tilde{\boldsymbol{\gamma}}_2$,
whose ``hyper-diagonal" $(\vr_1=\vr_1',\vr_2=\vr_2')$ provides a {\it bona‑fide} two‑qubit state:
\begin{align}
\bL(\vr_1,\vr_2)
=\frac{\tilde{\boldsymbol{\gamma}}_2(\vr_1,\vr_2)}
{\mathrm{Tr}\,[\tilde{\boldsymbol{\gamma}}_2(\vr_1,\vr_2)]},
\label{eqn:D}
\end{align}
where $\mathrm{Tr}$ denotes a trace in the two‑spin space. This step involves a non‑empirical 
{trace normalization} and yields a proper density operator (unit trace, positive semidefinite). In systems with spin‑rotational symmetry, $\bL$ assumes the Werner form \cite{Werner89}
\begin{equation}
\bL(\vr_1,\vr_2)
= p(\vr_1,\vr_2)\,|\Psi^{-}\rangle\langle\Psi^{-}|
+\bigl[1-p(\vr_1,\vr_2)\bigr]\;\mathcal{I}/4.
\label{werner}
\end{equation}
Here, $|\Psi^{-}\rangle$ is the singlet state, $\mathcal{I}$ is the two‑spin identity, $p(\vr_1,\vr_2) =  |\gamma_1(\vr_1,\vr_2)|^2 / \left[2 n (\vr_1)n(\vr_2)- |\gamma_1(\vr_1,\vr_2)|^2 \right]$ is the excess singlet probability determined by the spin coherence in $\tilde{\boldsymbol{\gamma}}_2$, and $n(\vr)=\gamma_1(\vr,\vr)$ is the electron density.

We quantify the spin entanglement encoded in $\bL$ through the concurrence,
\begin{align}
\tcalC(\vr_1,\vr_2)
=\max\!\left\{0,\;2\tlambda_{\max}(\vr_1,\vr_2)-\sum_{i=1}^4\tlambda_i(\vr_1,\vr_2)\right\},
\label{eqn:C1}
\end{align}
where $\{\tlambda_i\}$ are the eigenvalues of the Wootters matrix $R=\sqrt{\sqrt{\bL}\,{\bL_{\rm SF}}\,\sqrt{\bL}}$, with spin-flipped (SF) density matrix ${\bL_{\rm SF}}=(\sigma_y\!\otimes\!\sigma_y)\bL^*(\sigma_y\!\otimes\!\sigma_y)$~\cite{Wootters1998}. For the states considered here ${\bL_{\rm SF}}=\bL$, so that $R=\bL$ and $\{\tlambda_i\}$ reduce to the eigenvalues of $\bL$ itself. For a Werner state this gives
\begin{equation}
\tcalC(\vr_1,\vr_2)=\max\!\left\{\frac{3p(\vr_1,\vr_2)-1}{2},\,0\right\},
\label{conc}
\end{equation}
so that $\tcalC=0$ for $p\le 1/3$ and $\tcalC=1$ when $\bL$ coincides with a pure singlet state ($p=1$). The level at which $\gamma_1$ is determined (e.g., RHF, or beyond) propagates through Eqs.~\eqref{eqn:dm2dm1}–\eqref{conc} and thus influences the degree of entanglement given by $\tilde\calC$.
Again, the notation underlines that  the concurrence computed for estimating the localization 
uses $\tgm$ [see Eq. \eqref{eqn:dm2dm1}] rather than the pure-state representable $\gamma_2$.
{Because $\tgm$ is generally not pure-state 
$N$-representable, $\bL$ naturally incorporates singlet–triplet mixing that emerges in strongly correlated regimes.}

{This construction has three immediate consequences: (i) it associates maximum localization with singlet configurations rather than with fermions endowed with a bosonic kinetic-energy form \cite{Savin92,Savin94,Savin96} — an interpretation which, if taken too far, would contradict the fermionic origin of chemical bonding, underscoring that full nonlocality is conceptually essential; (ii) consistent with this, it predicts $\tcalC = 0$
whenever the exchange term is small compared to the Hartree term in Eq. \eqref{eqn:dm2dm1} [see Eq. \eqref{conc}]; and (iii) $\tcalC$ is rigorously bounded in $[0,1]$
with a direct quantum-information meaning, requiring neither heuristic normalization nor {\it ad hoc} non-linear remapping (see below).}  

 \begin{figure}[!t]
    \centering
    \includegraphics[width=\columnwidth]{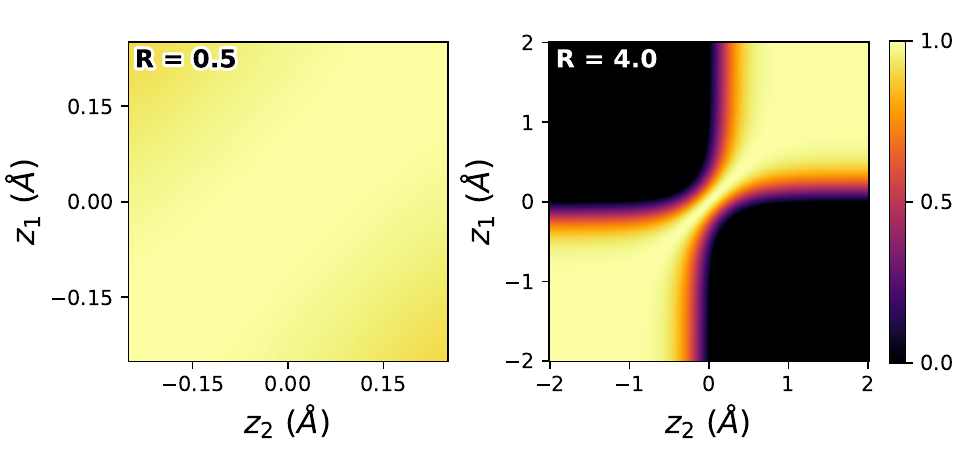}
    \caption{$\tcalC(z_1,z_2)$ for H$_2$ at the CASSCF(2,2) level. The coordinates $z_1$ and $z_2$ are along the internuclear axis so that nuclei are at corners.
Positions $z_1$ and $z_2$ are sampled along the internuclear axis between the nuclei.
Left: {compressed geometry} ($R=0.5$ \AA).
Right: stretched configuration ($R=4.0$ \AA).}
    \label{fig1}
\end{figure}
Increasing internuclear separation in molecules enhances static correlation and leads to a gradual disentanglement of the
spin pairs (see below).

\begin{figure}[!b]
    \centering
    \includegraphics[width=\columnwidth]{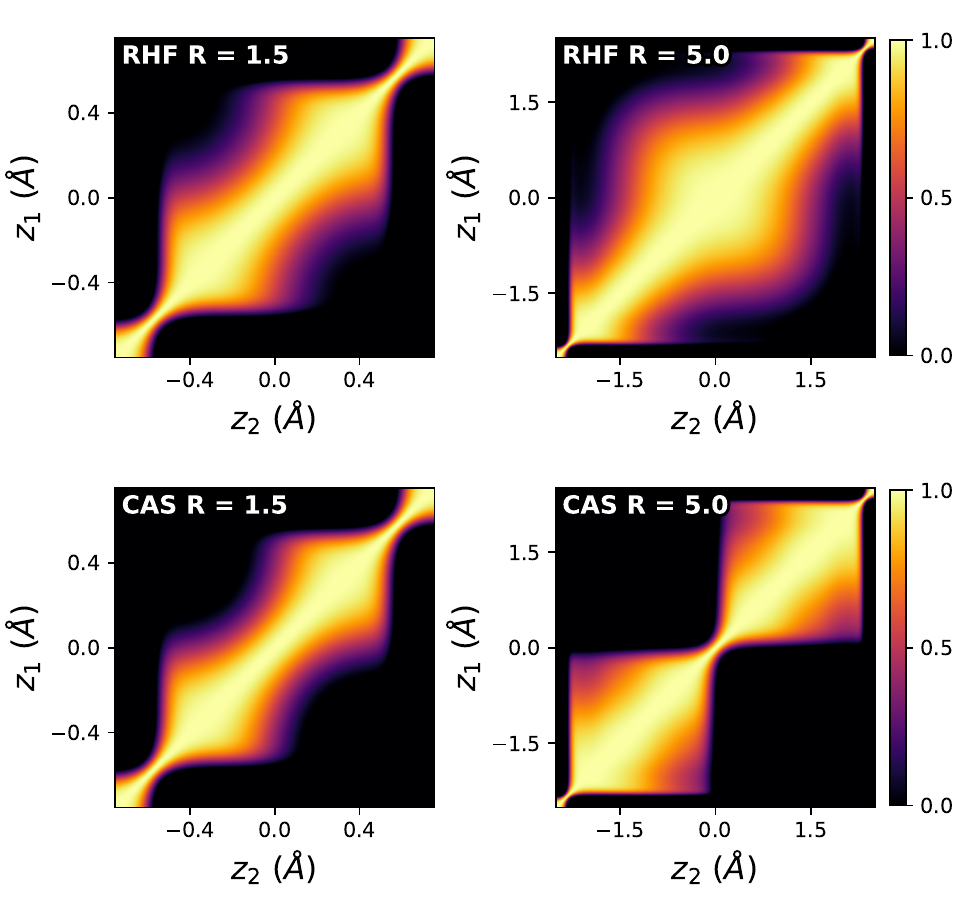}
    \caption{$\tcalC(z_1,z_2)$ for F$_2$. The coordinates $z_1$ and $z_2$ are along the internuclear axis so that nuclei are at corners. Top: RHF; bottom: CASSCF(2,2). Left: near the equilibrium geometry ($R=1.5$ \AA). Right: stretched configuration ($R=5.0$ \AA).}
    \label{fig2}
\end{figure}

Next, it is instructive to see how the conventional ELF is related to $\tilde\calC$.
To recover the ELF from our formulation, four reductions must be imposed:
(i) One first replaces $\tilde{\boldsymbol{\gamma}}_2(\vr,\vr+\vu)$ with its angular average over $\vu$, 
thereby discarding directional information;
(ii) one then expands the  concurrence for small inter-electronic separations $u$, keeping the leading term,
$\tcalC(\vr,u)
\approx \max\!\left\{0,\, 1 - \left[u/l_E(\vr)\right]^2 + \ldots\right\},
$
where (assuming real-valued natural orbitals)
$
l_E(\vr)
\propto
\left\{
\tau(\vr)/n(\vr)
-|\nabla n(\vr)|^2/4n^2(\vr)
\right\}^{-1/2}
$
is a characteristic entanglement length, {expressed in terms of the particle density
$n(\vr) = \sum_i n_i |\psi_i(\vr)|^2$
and semi-local quantities: the kinetic-energy density
$\tau(\vr) = \nabla \cdot \nabla' \gamma_1(\vr,\vr')|_{\vr'=\vr} = \sum_i n_i |\nabla \psi_i(\vr)|^2$
and the density gradient $\nabla n(\vr)$};
(iii) one may normalize $l_E(\vr)$ to its homogeneous‑electron‑gas value $l_E^{\rm unif}
(\vr)\propto n(\vr)^{-1/3}$;
and (iv) finally, one may apply a nonlinear logistic‑like mapping that is not dictated by any known physical principle,
$ {\rm ELF}(\vr)
= 1 / \,\{1 + [\,l_E^{\rm unif}(\vr)/l_E(\vr)\,]^2\}\,$,
which reproduces the Becke–Edgecombe ELF when $\tgm$ is evaluated at the RHF level. 
By contrast, the concurrence itself retains full two-point information and requires {\it no} heuristic normalization or empirical logistic‑like mapping.

\begin{figure}[!b]
    \centering
    \includegraphics[width=\columnwidth]{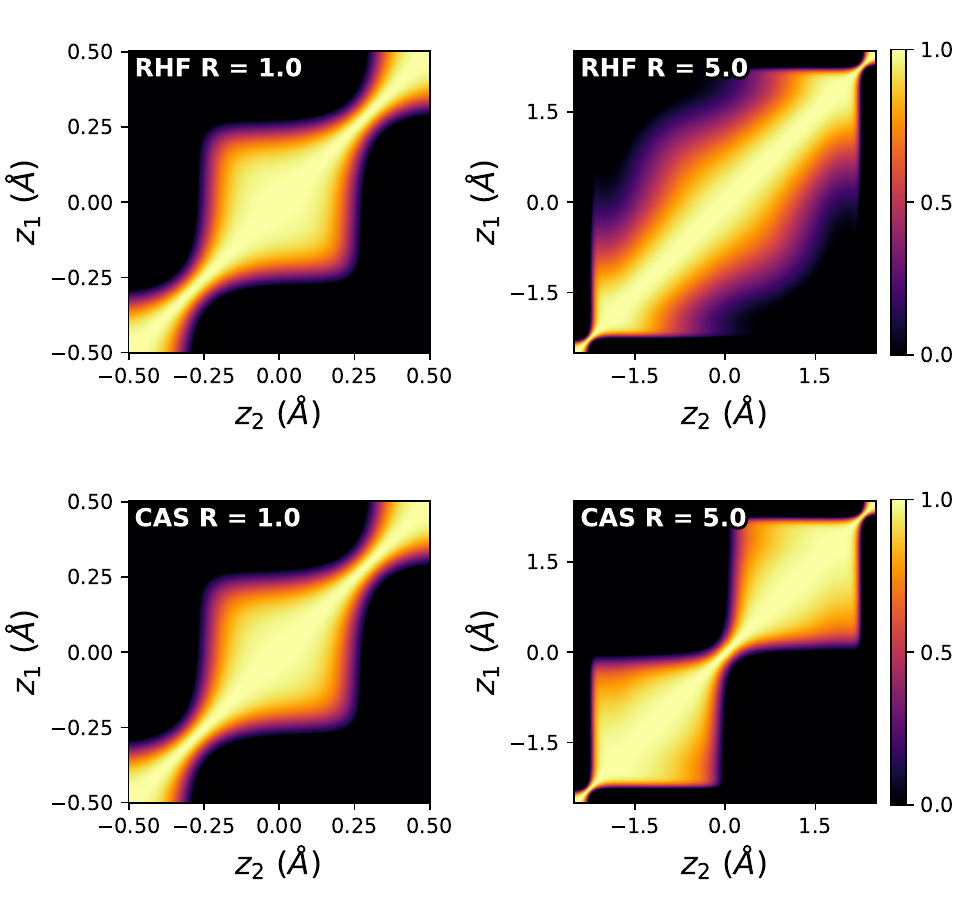}
    \caption{$\tcalC(z_1,z_2)$ for N$_2$. The coordinates $z_1$ and $z_2$ are along the internuclear axis so that nuclei are at corners.
Top: RHF; bottom: CASSCF(6,6). Left: near equilibrium ($R=1.0$ \AA). Right: stretched geometry ($R=5.0$ \AA). 
Sampling along the internuclear axis.}
    \label{fig3}
\end{figure}

The above analysis provides a unified perspective to view both the traditional RHF‑based ELF and its correlated extensions~\cite{Savin96,Kohout2004,Feixas2010} as different instances of the same reduction. Furthermore, it highlights the empirical elements in the previous ELF constructions. 

{We emphasize that the reduction from $\tcalC(\vec{r}_1,\vec{r}_2)$ to the ${\rm ELF}(\vr)$ yields $l_E(\vec{r})$, a natural three-dimensional indicator that assigns a quantum-information meaning to a key ingredient of the ELF—the Pauli kinetic energy excess (here divided by the particle density). This observation was already made at the HF/Kohn–Sham level in Ref.~\cite{Pittalis15a}; the present work extends it to correlated 1RDMs and identifies the full six-dimensional $\tcalC(\vec{r}_1,\vec{r}_2)$ as the underlying nonlocal parent quantity encoding electron localization from the outset.
Conversely, this perspective also suggests that $\tcalC(\vec{r}_1,\vec{r}_2)$ can serve as a starting point for constructing other, quantum-information based, lower-dimensional projections.}

\begin{figure}[!t]
    \centering
    \includegraphics[width=\columnwidth]{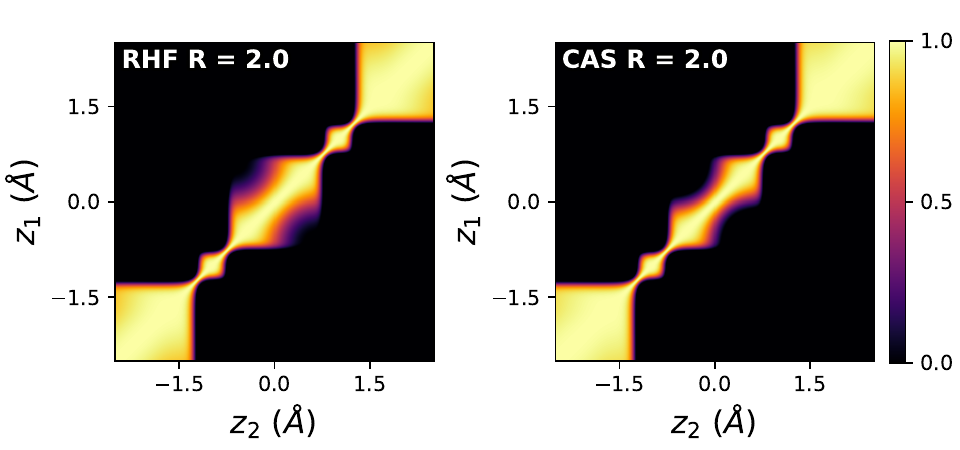}
    \caption{$\tcalC(z_1,z_2)$ for N$_2$ at $R=2$ \AA. Nuclei are at $z=-1$ and $z=1$ \AA. 
Sampling extends across and beyond the atomic regions
Left: RHF; right: CASSCF(6,6).}
    \label{fig4}
\end{figure}

{\it Applications.}~
We examine the dissociation of closed‑shell diatomic molecules into open‑shell atomic fragments, including both covalent and ionic bonds as well as a representative singlet excitation. We perform RHF and complete active space self-consistent field (CASSCF) \cite{BookCAS} calculations \footnote{We employ \texttt{pyscf} \cite{pyscf2018} and post‑processed the output with custom scripts. All systems are treated in the aug-cc-pvtz basis set, except the hydrogen molecule, for which the cc-pvtz basis is used.}. In the following, CASSCF($n$,$m$) indicates a CASSCF calculation with $n$ active electrons distributed within $m$ active orbitals.

All figures show a projection $\tcalC(z_1,z_2)$ onto the internuclear $z$‑axis, with the two atoms appearing in opposite corners of each panel, except in Fig.~\ref{fig4} where a wider window is shown.
Each value $\tcalC(z_1,z_2)$ gives the entanglement of the two‑spin state extracted from $\tgm$ [see Eq.~\eqref{eqn:D}].  
Regions with $\tcalC>0$ contain entangled pairs, with larger values indicating stronger singlet character.  
Along the diagonal ($z_1=z_2$), antisymmetry enforces $\tcalC=1$, while along the anti‑diagonal ($z_1=-z_2$) the electrons occupy opposite sides of the bond midpoint.

The ground state of H$_2$ is a pure singlet with only two electrons. Hence, the concurrence computed directly from the corresponding pure‑state 2RDM equals one everywhere, independently of whether the underlying wavefunction is RHF, CASSCF(2,2), or exact.  
Thus, a visualization based solely on $\gamma_2$ cannot reveal whether an approximate wavefunction correctly describes dissociation.

Figure~\ref{fig1} shows that this limitation is resolved by resorting to the auxiliary $\tgm$.  
At the RHF level, $\tcalC$ remains equal to one, consistent with the fact that RHF does not describe bond breaking.  
At the CASSCF level, however, triplet components mix in as the bond is stretched, and $\tcalC$ decreases accordingly.  
Near equilibrium (left), the entanglement is only mildly reduced upon moving the electrons toward opposite atoms; beyond the Coulson–Fischer point {\cite{Couls49}} (right), the collapse of $\tcalC$ clearly renders the idea of a stretched bond and the emergence of two separate fragments.

For multi‑electron systems, F$_2$ in Fig. \ref{fig2} and N$_2$ in Fig. \ref{fig3}, the concurrence maps encode both bonding and shells.  
Bonding information resides in the regions where $z_1$ and $z_2$ lie on opposite sides of the diagonal: a central bright island connecting the two atoms indicates inter‑atomic singlet entanglement, while its collapse into a thin bridge (a size to be intended relative to the internuclear distance) signals dissociation. Thus, the effect of static correlations as compared to a mean-field solution is made apparent. We notice that 
the narrow dip at the equilibrium distance, previously reported in ELF studies of F$_2$ \cite{Scemama2004},
is also visible in the fully non‑local concurrence.
Bright rectangular blocks close to each nucleus---where both electrons reside on the same atom---reflect the atomic shell structure: within a given shell, $\tcalC$ remains high, whereas dark channels mark the shell boundaries (see Figs.~\ref{fig2} and \ref{fig3}).

Energetically, RHF is an acceptable approximation near equilibrium but fails in stretched geometries, where static correlation becomes essential.
Consistently, at the RHF level the concurrence spuriously increases away from the bond center in stretched regimes, artificially suggesting a reinforcement of the bond (upper right panel of Figs.~\ref{fig2} and \ref{fig3}).
At the CASSCF level, by contrast, the collapse of $\tcalC$ between the bright atomic blocks (bottom right panel) is the real‑space signature of bond breaking and the emergence of two independent fragments.

Figure~\ref{fig4} provides a zoomed‑out visualization for N$_2$ in a slightly stretched geometry, making lone pairs visible through $\tcalC$. It is also clear that, at this distance, correlation effects are minor, and yet not entirely negligible along the stretched bond.

\begin{figure}[htbp]
    \centering
    \includegraphics[width=\columnwidth]{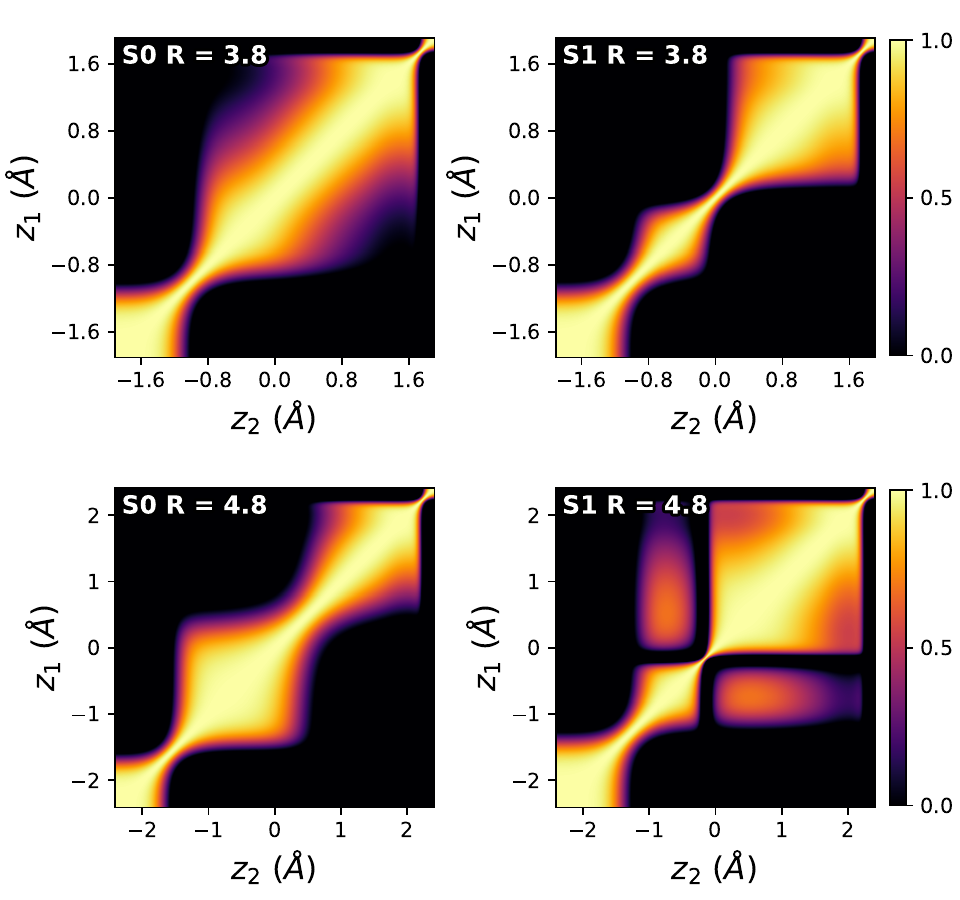}
    \caption{$\tcalC(z_1,z_2)$ for LiF from SA‑CASSCF(2,2). The coordinates $z_1$ and $z_2$ are along the internuclear axis so that nuclei are at corners (Li on the left, F on the right).
Left: ground state $S_0$. Right: first excited singlet $S_1$.
Top: $R=3.8$ \AA; bottom: $R=4.8$ \AA.
$z_1$ and $z_2$ are sampled along the internuclear axis.
The avoided crossing occurs at SA-CASSCF(2,2) near $R\approx 4.3$ \AA.}
    \label{fig5}
\end{figure}

We next examine LiF, which passing through an avoided crossing undergoes an ionic–to–neutral charge transfer upon dissociation—a regime beyond the capabilities of RHF.
Equal‑weight SA‑CASSCF(2,2) calculations for the $S_0$ and $S_1$ singlets correctly resolve the avoided crossing near $R\approx 4.3$~\AA{}.
Figure~\ref{fig5} shows $\tcalC$ at internuclear distances about $0.5$~\AA{} before and $0.5$~\AA{} after this crossing.
Before the crossing (top row), $S_0$ is ionic, with $\tcalC$ displaying two disconnected blocks centered on the atoms, while
$S_1$ is covalent, featuring a bright diagonal island ($z_1\!\approx\! z_2$) between the atomic regions.
After the crossing (bottom row), $S_0$ becomes covalent and its $\tcalC$ develops a clear bridging island, while $S_1$ becomes ionic and its $\tcalC$ shows separated blocks that merge into a broader region.
The key diagnostic feature is thus the presence (covalent) or absence (ionic) of a distinct diagonal bridging island; after the crossing, the covalent island shrinks and merges into the wider ionic pattern.

{\it Conclusions.}~ We have introduced a quantum-information-based measure of electron localization derived from the spin entanglement encoded in a normalized two‑spin state built from the $\gamma_1$-derived Fock--Dirac two-body object. By extracting a spatially resolved two‑spin state and evaluating its concurrence, the method captures both local and non‑local aspects of electronic structure, dispensing with the short‑range assumptions and empirical elements of traditional localization functions. This establishes a rigorous quantum‑information foundation for next‑generation localization indicators relevant to both density‑functional and wave‑function methodologies.

Notably, the calculation involves the one‑body reduced density matrix while still reflecting non‑local correlations. This feature enables a computationally efficient measure of quantum correlations that can be evaluated with existing software and readily integrated into data-driven workflows \cite{ShaoPavanello2023,MartinezPavanello2026} for enhanced electronic‑structure modeling.

{\it Acknowledgments.}~
The authors acknowledge financial support from: the Ministero dell’Universit\`a e della Ricerca (MUR) under the Project PRIN 2022 number 2022W9W423 (SP, FT, CA) and the PNRR Project PE0000023-NQSTI (FT);
an Australian Research Council (ARC) Discovery Project DP200100033 (TG) and an ARC Future Fellowship FT210100663 (TG).

\setlength{\bibsep}{0pt}
\vspace{-10pt}


\begin{thebibliography}{52}
\bibitem [{fno()}]{fnote_semi}%
  \BibitemOpen
  \href@noop {} {}\bibinfo {note} {In standard DFT parlance, a form that
  depends on quantities at a given point and on its neighborhood through
  first-order spatial derivatives of the natural orbitals.}\BibitemShut {Stop}%
\end{thebibliography}
\end{document}